\documentclass[twocolumn,preprint,pra,showpacs,showkeys]{revtex4-1}
\usepackage{graphicx,amsmath,amsfonts,amssymb}
\usepackage{times}
\begin{document}

\title{ Quantum discord dynamical behaviors due to initial system-cavity correlations}
\author{Ying-Jie Zhang,$^{1}$ Xu-Bo Zou,$^{2}$ Yun-Jie Xia,$^{1,}$}
\thanks{Email: yjxia@mail.qfnu.edu.cn}

\affiliation {$^{1}$Shandong Provincial Key Laboratory of Laser
Polarization and Information Technology, Department of Physics, Qufu
Normal
University, Qufu 273165, People's Republic of China\\
 $^{2}$Key Laboratory of Quantum Information, Department of Physics, University
of Science and technology of China, Hefei 230026, People's Republic
of China }
\date{\today}

\begin{abstract}
  We analyze the roles of initial correlations between
  the two-qubit system and a dissipative cavity on quantum discord
  dynamics of two qubits. Considering two
initial system-cavity states, we show that the initial system-cavity
correlations not only can initially increase the two-qubit quantum
discord but also would lead to a larger long-time quantum discord
asymptotic value. Moreover, quantum discord due to initial
correlations is more robust than the case of the initial factorized
state. Finally, we show the initial correlations' importance for
dynamics behaviors of mutual information and classical correlation.
\end{abstract}


\maketitle

\noindent{\it{Introduction}}: Recently, lots of interest have been
devoted to the definition and understanding of correlations in the
quantum systems. Entanglement is a kind of quantum correlation that
has been playing a central role in quantum information and
communication theory [1]. However there are other nonclassical
correlations apart from entanglement [2-4] that can be of great
importance to these fields. In order to characterize all
nonclassical correlations, Ollivier and Zurek introduced what they
called quantum discord. They marked the beginning of a new line of
research shifting the attention from the entanglement vs.
separability dichotomy to the quantum vs. classical paradigm. The
quantum discord measures quantum correlations of a more general type
than entanglement, there exists separable mixed states having
nonzero discord [5]. Interestingly, it has been proven both
theoretically and experimentally that such states provide
computational speedup compared to classical states in some quantum
computation models [5,6].\\
 \indent Understanding of the quantum dynamics of open systems is a
 very important task in many areas of physics ranging from quantum
 optics to quantum information processing and to quantum cosmology.
 Therefore, the study of quantum and classical correlations'
 dynamical behaviors in the presence of both Markovian [7,8] and
 non-Markovian [9,10] decoherence has attracted much attention in recent
 years. It is believed that the quantum correlations measured by the
 quantum discord, in the Markovian case, decay exponentially in
 time and vanish only asymptotically [11], contrary to the
 entanglement dynamics where sudden death may occur [12]. In
 particular, Refs. [13,14] have discovered that quantum discord can
 be completely unaffected by Markovian depolarizing channels or non-Markovian depolarizing channels for long intervals of
 time, and this phenomenon has been observed experimentally [15]. In these above studies, quantum discord dynamics of open
 system in which system and environment are initially separable has
 been analyzed in detail. As everyone knows it is possible, and
 often unavoidable in experiment, to create correlations between
 system and environment, therefore such correlations will play
 important roles in the time evolution of the system. So the study
 of quantum correlation dynamics due to the initial
 system-environment correlations is certainly necessary.\\
 \indent The influence of initial correlations on the open system dynamics has recently been intensively studied [16-22].
  In order to clear what happens to the quantum discord in this situation, we study an exactly solvable model for the time
 evolution of two atoms interacting with a lossy cavity. In this model where the coupling strength effects are considerable,
the presence of system-cavity correlations invalidates the initial
state in which system and cavity are independent. We investigate
quantum discord dynamics for two initial states of two atoms and the
lossy cavity, in which the initial reduced density matrixes of the
atomic system and the cavity are the same. Under a certain
condition, the initial system-cavity correlations not only can
initially increase the atomic quantum discord but also would lead to
a larger long-time quantum discord asymptotic value. For another
condition, correlations between two atoms and the lossy cavity, can
more effectively restrain the reduction of quantum discord than the
case of the factorized state. And then we also analyze the different
dynamics behaviors of mutual information and classical correlation
due to the initial system-cavity correlations. These findings
obtained from our article show that, dynamics of mutual information,
quantum discord and classical correlation not only depend on the
system degrees of freedom,
the initial system-cavity correlations must also be properly taken into account.\\
 \noindent{\it{Quantum discord and classical correlation}}: We now
present a brief review of the classical correlation and quantum
discord. In classical information theory, the information can be
quantified by Shannon entropy
$H(X)=-\sum_{x}P_{|X=x}{\log}P_{|X=x}$, where $P_{|X=x}$ is the
probability with $X$ being $x$. Similarly, the joint entropy, which
measures the total uncertainty of a pair of random variables $X$ and
$Y$, is defined as
$H(X,Y)=-\sum_{x,y}P_{|X=x,Y=y}{\log}P_{|X=x,Y=y}$, with
$P_{|X=x,Y=y}$ being the probability in the case of $X=x$ and $Y=y$.
Then the total correlation between $X$ and $Y$ can be measured by
the mutual information which is defined as
$I(X:Y)=H(X)+H(Y)-H(X,Y)$, whose quantum version can be written as
[23]
\begin{equation}
\mathcal {I}(X:Y)=S(\rho_{X})+S(\rho_{Y})-S(\rho_{XY}),\label{01}
\end{equation}
where $S(\rho)=-Tr(\rho\log\rho)$ is the von Neumann entropy of
$\rho$, and $\rho_{X}(\rho_{Y})$ is the reduced density matrix of
$\rho_{XY}$ by tracing out $Y(X)$. By introducing the conditional
entropy $H(X|Y)=H(X,Y)-H(Y)$, we can rewrite the mutual information
as
\begin{equation}
I(X:Y)=H(X)-H(X|Y),\label{02}
\end{equation}
where
$H(X|Y)=\sum_{y}p_{Y=y}H(X|Y=y)=-\sum_{x,y}p_{X=x,Y=y}{\log}p_{X=x,Y=y}$
is the conditional entropy of the random variables $X$ and $Y$ for
the average uncertainty about the value of $X$ given that the value
of $Y$ is known. In order to generalize the above equation to the
quantum domain, we measure the subsystem $Y$ by a complete set of
projectors $\prod_{i}$, corresponding to the outcome $i$, which
yields $\rho_{X|i}=Tr_{Y}(\prod_{i}\rho_{XY}\prod_{i})/p_{i}$, with
$p_{i}=Tr_{XY}(\prod_{i}\rho_{XY}\prod_{i})$. Then the quantum
mutual information can alternatively defined by
\begin{equation}
\mathcal
{J}_{\prod_{i}}(X:Y)=S(\rho_{X})-S_{\prod_{i}}(X|Y),\label{03}
\end{equation}
where $S_{\prod_{i}}(X|Y)=\sum_{i}p_{i}S(\rho_{X|i})$ is conditional
entropy of the quantum state. The above quantity strongly depends on
the choice of the measurements $\{\prod_{i}\}$. By maximizing
$\mathcal{J}_{\prod_{i}}(X:Y)$ over all $\{\prod_{i}\}$, we define
the classical correlation between $X$ and $Y$
\begin{equation}
\mathcal {C}(X:Y)=\max_{\prod_{i}}{J}_{\prod_{i}}(X:Y),\label{04}
\end{equation}
and the quantum discord as
\begin{equation}
\mathcal {D}(X:Y)=\mathcal {I}(X:Y)-\mathcal {C}(X:Y),\label{05}
\end{equation}
which is interpreted as a measure of the quantum correlation [2-4].
It is zero only for states with classical correlations and nonzero
for states with quantum correlations. In particular, quantum discord
is equal to the entanglement of formation for pure states, it is not
true for mixed states, since some states present finite quantum
discord even without entanglement [2]. \\
\noindent{\it{The model}}: We consider two atoms $A$ and $B$
interacting with a dissipative cavity. The Hamiltonian of such total
system in the rotating-wave approximation is given by
$H=H_{0}+H_{int}$, which, in the basis
$\{|gg\rangle,|eg\rangle,|ge\rangle,|ee\rangle\}$ reads
\begin{equation}
H_{0}=\omega_{0}(\sigma^{A}_{+}\sigma^{A}_{-}+\sigma^{B}_{+}\sigma^{B}_{-})+\omega_{c}a^{\dag}a,\label{06}
\end{equation}
\begin{equation}
H_{int}=\Omega(\sigma^{A}_{+}a+\sigma^{B}_{+}a)+h.c.,\label{07}
\end{equation}
here $\sigma^{A}_{\pm}$, $\sigma^{B}_{\pm}$ are, respectively, the
Pauli raising and lowering operators for atoms $A$ and
 $B$, $\omega_{0}$ is the Bohr frequency of two atoms, $a$ and $a^{\dag}$ are the annihilation and creation operators
 for the cavity mode, which is characterized by the
 frequency $\omega_{c}$ and the coupling constant $\Omega$. For the
 sake of simplicity, in the following we assume that the two atoms
 interact resonantly with the dissipative cavity mode.
 \\
\indent In this section, we investigate the dynamics of the two
atoms interacting with a dissipative cavity by making use of the
master equation. Firstly, we focus on the case in which the total
system contains only one excitation. In the interaction picture, the
exact dynamics of the two atoms is contained in the following master
equation
\begin{eqnarray}
\frac{d\rho}{dt}=&-&i[H_{int},\rho]-\frac{\Gamma}{2}[a^{\dag}a\rho-2a{\rho}a^{\dag}+{\rho}a^{\dag}a],\label{08}
\end{eqnarray}
where $\rho$ is the density operator for the two atoms and the
cavity mode, $\Gamma$ is the decay rate of the cavity mode. In order
to find the dynamics of two atoms, we solve the master equation in
Eq. (8). In the case which the total system contains only one
excitation, we need to solve a set of $16$ differential equations
obtained from Eq. (8). Then, tracing out the cavity degree of
freedom, we obtain the reduced density matrix of the atomic system
for the total system
which only contains one excitation. \\
\indent Then we treat another case that the total system contains at
most two excitations. In this case the dynamics of two atoms can be
effectively described by a four-state system in which three states
are coupled to the cavity mode in a ladder configuration, and one
state is completely decoupled from the other states and form the
field. In the basis
$\{|0\rangle=|gg\rangle,|+\rangle=(|eg\rangle+|ge\rangle)/\sqrt{2},|-\rangle=(|eg\rangle-|ge\rangle)/\sqrt{2},|2\rangle=|ee\rangle\}$,
the Hamiltonians (6) and (7) can be rewritten
\begin{equation}
H'_{0}=2\omega_{0}|2\rangle\langle2|+\omega_{0}(|+\rangle\langle+|+|-\rangle\langle-|)+\sum_{k}\omega_{k}a^{\dag}_{k}a_{k},\label{10}
\end{equation}
\begin{equation}
H'_{int}=\sqrt{2}\Omega(|+\rangle\langle0|a+|2\rangle\langle+|a)+h.c..\label{11}
\end{equation}
From the total Hamiltonian given by (\ref{10}) and (\ref{11}), the
subradiant state $|-\rangle$ does not decay, and the superradiant
state is coupled to states $|0\rangle$ and $|2\rangle$ via the
cavity mode. The transitions $|0\rangle\rightarrow|+\rangle$ and
$|+\rangle\rightarrow|2\rangle$ have the same frequencies and
identically coupled with the cavity. In the interaction picture, the
dynamics of two atoms interacting with a lossy cavity can be treated
in the following master equation
\begin{eqnarray}
\frac{d\rho}{dt}=&-&i[H'_{int},\rho]-\frac{\Gamma}{2}[a^{\dag}a\rho-2a{\rho}a^{\dag}+{\rho}a^{\dag}a].\label{12}
\end{eqnarray}
Focusing on the case in which there are at most two excitations in
the total system, we should solve a set of $64$ differential
equations obtained from Eq. (12). The symmetry properties of
the system allow to transform the set of differential equations into
decoupled subset of differential equations of smaller size. Finally,
we obtain the density matrix of the reduced atomic
system by tracing out the cavity degree of freedom. \\
\indent To illustrate the roles of the initial correlations between
two atoms and cavity on the quantum discord dynamics of two atoms,
we consider a initial condition
$\rho^{(1)}_{ABc}=|\Psi\rangle_{ABc}\langle\Psi|$, with
$|\Psi\rangle_{ABc}=\alpha|ge0\rangle+\beta|eg0\rangle+\gamma|gg1\rangle$
(having correlations), and here
$|\alpha|^{2}+|\beta|^{2}+|\gamma|^{2}=1$
($\alpha,\beta,\gamma\neq0$), $|e\rangle$ and $|g\rangle$ are the
exited state and ground state of atoms, $|0\rangle$ and $|1\rangle$
are the vacuum state and the single-photon state of the lossy
cavity.  Obviously, two atoms and the cavity in this initial state
is
\begin{equation} \rho_{AB}=\left(
       \begin{array}{cccc}
         |\gamma|^{2} & 0 & 0 & 0 \\
         0 & |\alpha|^{2} & \alpha\beta^{*} & 0 \\
         0 & \alpha^{*}\beta & |\beta|^{2} & 0 \\
        0 & 0 & 0 & 0\\
       \end{array}
     \right),\label{13}
\end{equation}
\begin{equation} \rho_{c}=\left(
       \begin{array}{cccc}
          |\alpha|^{2}+|\beta|^{2}& 0  \\
         0 & |\gamma|^{2} \\
       \end{array}
     \right).\label{14}
\end{equation}
Therefore, the initial correlations must illustrate clear effects on
the dynamics of the atomic quantum discord. In the following, by
comparing to the second initial condition that the two-atom system
and the cavity are in the factorized state
$\rho^{(2)}_{ABc}=\rho_{AB}\otimes\rho_{c}$, we mainly study the
different atomic quantum discord dynamical behaviors due to these
two initial atoms-cavity states, which contain the identical states of subsystems.\\
\indent For $\rho^{(1)}_{ABc}$, it is composed by the basis
$|ge0\rangle$, $|eg0\rangle$ and $|gg1\rangle$, that is to say the
initial correlation state $\rho^{(1)}_{ABc}$ satisfies the first
case in which the total system contains only one excitation. So the
density matrix of the total system at time $t$ can be calculated by
Eq. (8). However, not only the basis ($|ge0\rangle$,
$|eg0\rangle$ and $|gg1\rangle$) but also those basis ($|ge1\rangle$
and $|eg1\rangle$) are contained in the factorized state, so
$\rho^{(2)}_{ABc}$ actually consists of two parts, one satisfying
one excitation, and the other meeting two excitations. Hence the
evolutional density matrix of $\rho^{(2)}_{ABc}$ can be acquired by
Eqs. (8) and (12). Tracing out the cavity degree of
freedom, we obtain the density matrix of the reduced atomic system
for these two different initial states. In the basis
$\{|gg\rangle,|eg\rangle,|ge\rangle,|ee\rangle\}$, we measure the
atom $B$ from the matrix $\rho_{AB}(t)$ by projecting on
$\{\cos\vartheta|e\rangle_{B}+e^{i\phi}\sin\vartheta|g\rangle_{B},
e^{-i\phi}\sin\vartheta|e\rangle_{B}-\cos\vartheta|g\rangle_{B}\}$.
Then the quantum discord and classical correlation could be
calculated numerically using
Eqs. (4) and (5).\\
 \begin{figure}
\includegraphics[scale=0.8]{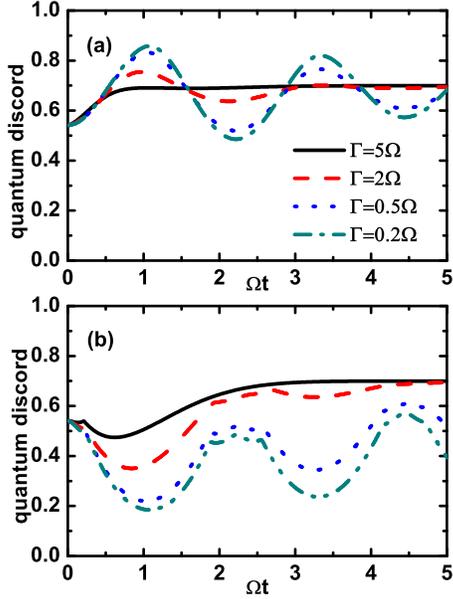}
\caption{\label{fig1}Time evolution of quantum discord of two atoms
as a function of the dimensionless quantity ${\Omega}t$ with
$\theta=\pi/3$ and $\varphi=3\pi/4$ (a) the initial system-cavity
correlated state $\rho^{(1)}_{ABc}$ as the total initial state, (b)
the factorized state $\rho^{(2)}_{ABc}$ as the total initial state.}
\end{figure}
 \noindent{\it{Numerical results and discussions}}: In this article we study a system
 whose dynamics is described by the well-known damped Tavis-Cummings model. To see the effects of initial system-cavity correlations explicitly, We consider two distinct
 initial states, the system-cavity correlated state $\rho^{(1)}_{ABc}$ and the factorized state $\rho^{(2)}_{ABc}$. The reduced density matrices
 for both the atomic system and the lossy cavity are the same. For simplicity,
 we choose the parameters $\alpha=\sin\theta\cos\varphi$, $\beta=\sin\theta\sin\varphi$ and $\gamma=\cos\theta$ ($\theta\in[0,\pi/2]$
 and $\varphi\in[0,\pi]$). In Fig.1, by choosing $\rho^{(1)}_{ABc}$ and $\rho^{(2)}_{ABc}$ as the initial states of the total system,
 we plot the time evolution of quantum discord for two atoms as a function of the dimensionless quantity ${\Omega}t$, with $\theta=\pi/3$ and $\varphi=3\pi/4$.
 We can observe that the atomic quantum discord can be initially increased to a maximum in the case $\rho^{(1)}_{ABc}$ (as shown in Fig.1(a)), which contains
 system-cavity correlations at $t=0$. While in the factorized initial state $\rho^{(2)}_{ABc}$, the initially decreased atomic quantum discord to a
 minimum would happen (as shown in Fig.1(b)). Hereafter, how much the
 coupling strength of our regime influences the quantum discord?
 Since $\Gamma$ is connected to the lossy cavity correlation time
 $\tau_{B}$ by the relation $\tau_{B}\approx1/\Gamma$, and $\Omega$
 is related to the time scale $\tau_{R}$ over which the state of the
 system changes, $\tau_{R}\approx1/\Omega$. So $\Gamma/\Omega<2$
 characterizes the strong coupling regime, and
 $\Gamma/\Omega>2$ means the weak coupling regime. Fig.1a presents
 the case for the initial system-cavity correlated state
 $\rho^{(1)}_{ABc}$, it can be seen that the initially increased quantum
 discord maximum increases with the decrease of the $\Gamma/\Omega$
 corresponding to the enhancement of the coupling strength. We also
 clearly see that for the factorized initial state
 $\rho^{(2)}_{ABc}$ in Fig.1b, the initially decreased quantum
 discord minimum decreases with the decrease of the $\Gamma/\Omega$.
\\
 \indent Then, by investigating the subsystem of two atoms, we shall compare the roles of different initial system-cavity
 states ($\rho^{(1)}_{ABc}$ and $\rho^{(2)}_{ABc}$) on the atomic quantum discord
 dynamics. Although the initial reduced density matrices for both the atomic system and the lossy cavity are
 the same as in the full calculation, Fig.2 shows that the presence of the system-cavity
correlations in the initial state changes the atomic quantum
 discord dynamics dramatically. In figs.2(a) and 2(b), quantum
 discord dynamics is given for the conditions $\theta=\pi/3$ and
 $\varphi=3\pi/4$ in the strong coupling regime and weak coupling
 regime, respectively. For the case $\rho^{(1)}_{ABc}$, the atomic
 quantum discord can firstly increase to a maximum, then
 periodically decrease to a long-time asymptotic value in the strong
 coupling regime. While in the weak coupling regime, quantum discord
 would reach a maximum at first and then gradually decrease to the
 asymptotic fixed value. For the other case $\rho^{(2)}_{ABc}$,
 quantum discord of two atoms initially reduce to a minimum, and
 eventually periodically increase to another long-time asymptotic
 value in the strong coupling regime (no oscillations are present in the weak coupling
 regime). Due to the subradiant state $|-\rangle$ in two initial
 states does not decay in the evolution process, the long-time
 asymptotic quantum discord would be acquired. The final long-time asymptotic
 value due to the system-cavity correlations is much larger than
 the case of the factorized state. So we can conclude that the
 initial system-cavity correlations not only can initially increase
 the atomic quantum discord but also would lead to a larger
 long-time quantum discord asymptotic value.\\
 \begin{figure}
\includegraphics[scale=0.85]{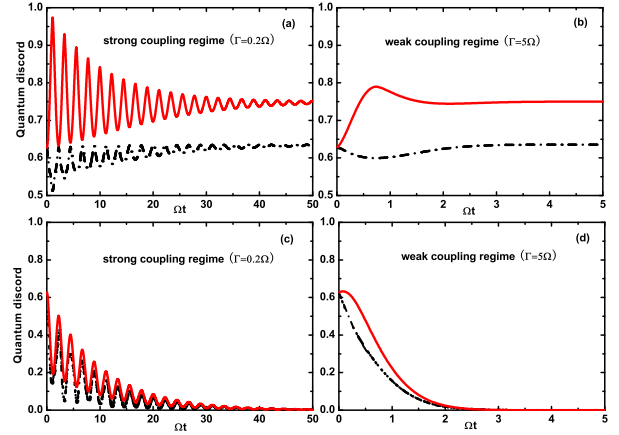}
\caption{\label{fig2}Time evolution of quantum discord of two atoms
as a function of the dimensionless quantity ${\Omega}t$ with
$\theta=\pi/3$, (a) $\varphi=3\pi/4$ and $\Gamma=0.2\Omega$, (b)
$\varphi=3\pi/4$ and $\Gamma=5\Omega$, (c) $\varphi=\pi/4$ and
$\Gamma=0.2\Omega$ and (d) $\varphi=\pi/4$ and $\Gamma=5\Omega$. For
the cases of (i) the initial system-cavity correlated state
$\rho^{(1)}_{ABc}$ ( red solid curve), (ii) the factorized state
$\rho^{(2)}_{ABc}$ (dark dash-dotted curve).}
\end{figure}
 \begin{figure}
\includegraphics[scale=0.8]{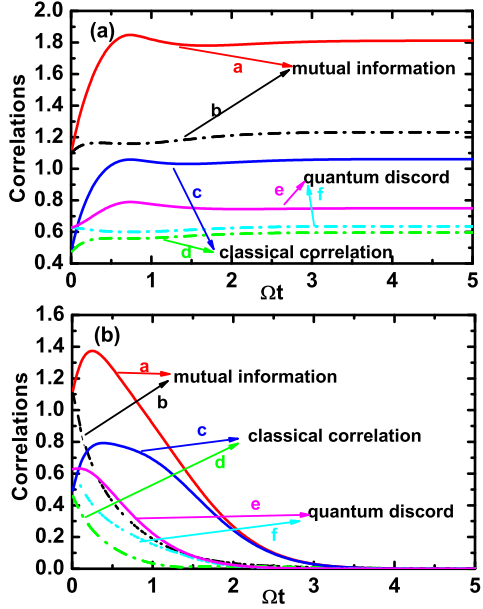}
\caption{\label{fig3}Correlation (mutual information, quantum
discord, classical correlation) dynamics of two atoms as a function
of the dimensionless quantity ${\Omega}t$ in the weak coupling
regime with $\theta=\pi/3$, (a) $\varphi=3\pi/4$ and
$\Gamma=5\Omega$, (b) $\varphi=\pi/4$ and $\Gamma=5\Omega$. For the
cases of (i) the initial system-cavity correlated state
$\rho^{(1)}_{ABc}$ ( the solid curves $a$, $c$ and $e$), (ii) the
factorized state $\rho^{(2)}_{ABc}$ ( the dash-dotted curves $b$,
$d$ and $f$).}
\end{figure}
\indent Figs.2(c) and 2(d) give the atomic quantum discord dynamics
for the conditions $\theta=\pi/3$ and
 $\varphi=\pi/4$ in the strong coupling regime and weak coupling
 regime, respectively. When $\varphi=\pi/4$, the initial atomic state does not
 contain the subradiant state $|-\rangle$, the asymptotic stationary state of two atoms is
 $\rho_{AB}(t\rightarrow\infty)=|gg\rangle{\langle}gg|$,
 so the long-time asymptotic
 value of the atomic quantum discord is zero. For both the initial
 system-cavity correlated state or the factorized state, in the strong coupling
 regime the atomic quantum discord presents damped oscillations while in the
 weak coupling regime quantum discord decays only asymptotically to zero. A comparison between the red solid curve and the dark
 dash-dotted curve in Figs.2(c) and 2(d) reveals that for the same initial atomic
state, the atomic quantum discord dynamics due to the initial
system-cavity
 correlations is more robust than the case of the factorized state. This suggests that correlations between two atoms and the lossy cavity,
 effectively restrains the reduction of atomic quantum discord.\\
\indent To further understand the roles of different initial
system-cavity states ($\rho^{(1)}_{ABc}$ and $\rho^{(2)}_{ABc}$) on
dynamics of the total correlations, we study the atomic mutual
information $\mathcal {I}$ and classical correlation $\mathcal {C}$,
respectively. Fig.3 shows the time evolution of the atomic mutual
information, quantum discord and classical correlation in the weak
coupling regime. It is clearly found that for the initial correlated
state $\rho^{(1)}_{ABc}$, the atomic classical correlation is
greater than quantum discord except in the initial period of time.
While the classical correlation is always less than quantum discord
for the initial factorized state $\rho^{(2)}_{ABc}$. When
$\varphi=3\pi/4$, because of the initial system-cavity correlations,
both the atomic classical correlation and mutual information can
firstly increase to a maximum, and then get to a fixed long-time
asymptotic value. In addition, the fixed long-time asymptotic values
of classical correlation and mutual information due to
$\rho^{(1)}_{ABc}$ are much larger than the case of
$\rho^{(2)}_{ABc}$ (Fig.3(a)). This result is same as quantum
discord dynamics. When $\varphi=\pi/4$, the atomic classical
correlation and mutual information due to the correlated state
$\rho^{(1)}_{ABc}$ would increase to a maximum value at first, then
reduce asymptotically to zero. This finding is different from
quantum discord dynamics, which only decays to zero (Fig.3(b)).
Comparing to the case of the initial factorized state
$\rho^{(2)}_{ABc}$, classical correlation and mutual information of two atoms also can be more robust.\\
\noindent{\it{Conclusion}}: In conclusion, we have studied quantum
discord of two qubits in the presence of initial system-cavity
correlations. The correlations between the qubits and a lossy cavity
is found to have important effect on the time evolution of quantum
discord when two qubits couple with a common lossy cavity. Finally,
we also analyze the different dynamics behaviors of mutual
information and classical correlation due to the initial
system-cavity correlations. In comparison with some recent work on
the initial factorized state between the system and environment, our
present work might be more practical to explain
the total correlation dynamics behaviors of the system.\\

 \noindent{\it{Acknowledgments}}: This work is supported by
National Natural Science Foundation of China under Grant Nos.
10774088 and 10947006, the Specialized Research Fund for the
Doctoral Program of Higher Education under Grant No. 20093705110001.

\end{document}